\documentclass[traditabstract]{aa}

\usepackage[colorlinks=true, linkcolor=blue, citecolor=blue, urlcolor=blue]{hyperref}
\usepackage{amsmath}
\usepackage{graphicx}
\usepackage{txfonts}
\usepackage{amssymb}
\usepackage[]{natbib}
\usepackage{mathrsfs}
\usepackage{caption}
\usepackage{float}
\usepackage{afterpage}
\usepackage{amstext}

\newcommand{\qt}[0]{\mbox{Qatar-1}}

\newcommand{\qtb}[0]{\mbox{Qatar-1b}}

\begin{document}

   \title{Testing connections between exo-atmospheres and their host stars}
   \subtitle{GEMINI-N/GMOS ground-based transmission spectrum of Qatar-1b}
   \author{C. von Essen$^1$, S. Cellone$^{2,3,4}$, M. Mallonn$^5$, S. Albrecht$^1$, R. Micul\'an$^{3,4}$, H.\,M. M\"uller$^6$} \authorrunning{C. von Essen et al. (2016)}
   \titlerunning{Transmission spectrum of \qtb}
   \offprints{cessen@phys.au.dk}

   \institute{$^1$Stellar Astrophysics Centre, Department of Physics and Astronomy, Aarhus University, Ny Munkegade 120, DK-8000 Aarhus C, Denmark\\ 
     $^2$Consejo Nacional de Investigaciones Cient\'{\i}ficas y T\'ecnicas, Godoy Cruz 2290, C1425FQB, Ciudad Aut\'onoma de Buenos Aires, Argentina\\
     $^3$Facultad de Ciencias Astron\'omicas y Geof\'{\i}sicas, Universidad Nacional de La Plata, Paseo del Bosque, B1900FWA, La Plata, Argentina\\
     $^4$Instituto de Astrof\'{\i}sica de La Plata (CCT-La Plata, CONICET-UNLP), Paseo del Bosque, B1900FWA, La Plata, Argentina\\
     $^5$Leibniz-Institut f\"ur Astrophysik Potsdam, An der Sternwarte 16, 14482, Potsdam, Germany\\
     $^6$Hamburger Sternwarte, Universit\"at Hamburg, Gojenbergsweg 112, 21029 Hamburg, Germany\\
              \email{cessen@phys.au.dk}
               }

   \date{Received January 27$^{th}$, 2017; accepted March 16$^{th}$, 2017}

\abstract{Till date, only a handful exo-atmospheres have been well
  characterized, mostly by means of the transit method. Some classic
  examples are \mbox{HD~209458b}, \mbox{HD~189733b}, \mbox{GJ-436b},
  and \mbox{GJ-1214b}. Data show exoplanet atmospheres to be
  diverse. However, this is based on a small number of cases. Here we
  focus our study on the exo-atmosphere of \qtb, an exoplanet that
  looks much like \mbox{HD~189733b} regarding its host star's activity
  level, their surface gravity, scale height, equilibrium temperature
  and transit parameters. Thus, our motivation relied on carrying out
  a comparative study of their atmospheres, and assess if these are
  regulated by their environment. In this work we present one primary
  transit of \qtb\ obtained during September, 2014, using the 8.1\,m
  GEMINI North telescope. The observations were performed using the
  GMOS-N instrument in multi-object spectroscopic mode. We collected
  fluxes of \qt\ and six more reference stars, covering the wavelength
  range between 460 and 746 nm. The achieved photometric precision of
  0.18 parts-per-thousand in the white light curve, at a cadence of
  165 seconds, makes this one of the most precise datasets obtained
  from the ground. We created 12 chromatic transit light curves that
  we computed by integrating fluxes in wavelength bins of different
  sizes, ranging between 3.5 and 20 nm. Although the data are of
  excellent quality, the wavelength coverage and the precision of the
  transmission spectrum are not sufficient to neither rule out or to
  favor classic atmospheric models. Nonetheless, simple statistical
  analysis favors the clear atmosphere scenario. A larger wavelength
  coverage or space-based data is required to characterize the
  constituents of \qtb's atmosphere and to compare it to the well
  known \mbox{HD~189733b}. On top of the similarities of the orbital
  and physical parameters of both exoplanets, from a long H$\alpha$
  photometric follow-up of \qt, presented in this work, we find
  \qt\ to be as active as \mbox{HD~189733}.}

\keywords{stars: planetary systems -- stars: individual: Qatar-1 --
  methods: observational}
          
   \maketitle

\section{Introduction}

Planetary transits present a unique opportunity to study the
properties of exoplanet atmospheres. During these events, a fraction
of the stellar light passes through the optically thin part of the
planetary atmosphere, picking up spectral features from it. The
technique that reveals the composition and extent of the atmosphere of
the planet is called transmission spectroscopy. To date, the chemical
composition has been characterized of some systems. Some classic
examples are \cite{Charbonneau2002} and \cite{Sing2008a}, who detected
atmospheric sodium in the atmosphere of \mbox{HD~209458b} using the
Hubble Space Telescope. This was further confirmed using high
resolution ground-based observations by \cite{Snellen2008}. From
near-infrared observations, \cite{Deming2013} added water to the
detected molecules. \cite{Redfield2008} detected sodium in the
atmosphere of \mbox{HD~189733b}, while \cite{Sing2011} found potassium
in the X0-2 system. \cite{Bean2010} carried out the first
characterization of the transmission spectrum of \mbox{GJ 1214b} from
the ground, and \cite{Gibson2013b} were pioneers in the
characterization of exo-atmospheres by means of multi-object
spectrographs. Although models predicted pressure-broadened absorption
features of the alkali metals in cloud-free atmospheres, observational
data collected at a high speed was evidencing something else: many hot
Jupiter spectra were best explained by clouds or hazes in the
planetary atmospheres \citep[e.g.,][]{Pont2013,Mallonn2016b}. A
complete overview of the exo-atmospheric zoo finally came from
\cite{Sing2016}, who pinpointed the richness of exo-atmospheric
composition in the first comparative analysis of hot Jupiters.

To maximize the chances to detect the exoplanetary atmosphere with
current ground-based instrumentation, the transiting systems need to
present two distinctive features: a large transit signal to reach the
signal-to-noise requirements, and low planetary surface gravity
(equivalently, a prominent scale height) to ease the identification of
the atmospheric signal \citep[see,
  e.g.,][]{LecavelierDesEtangs2008}. In the particular case of the
observing technique carried out in this work, reference stars of
similar brightness as the target are also needed within the field of
view. A transiting system offering all these conditions is
\qt. Transits in \qt\ \citep[\mbox{V$\sim$12.8};][]{droege_2006} were
first reported by \cite{Alsubai2011}, who characterized the host star
as an old \mbox{K-type} star, with \mbox{0.85~$M_\mathrm{\odot}$}, and
\mbox{0.82~$R_\mathrm{\odot}$}. Its hot Jupiter, \qtb, has a radius of
\mbox{$1.16$~$R_\mathrm{J}$} and orbits the star each $\sim$1.42 days
with an inclination angle of \mbox{$\sim 84^\circ$}. This geometry
implies nearly grazing transits. For \qt\ the transits are
\mbox{$\sim$2\%} deep, making them easy to detect and follow from the
ground. While the parent star is smaller and cooler than the Sun, the
climate of \qtb\ is extremely hot, reaching temperatures close to
\mbox{1400 K} \citep{Covino2013,Mislis2015}. The proximity between
exoplanet and host also means that the upper atmosphere of \qtb\ is
constantly battered by radiation whose strength is directly related to
the activity of its host. Atmospheric evaporation is thought to be
caused by ultraviolet and X-ray radiation from the parent star. Since
\qt\ is relatively far away \mbox{($\sim$200 pc)} an X-ray detection
and further characterization of the source requires prohibitively
large amounts of exposure time. Nonetheless, we and others have
characterized the activity of the star \citep[see][and Appendix
  \ref{Sec:Activity}]{Covino2013,Mislis2015}. We found that the
activity levels of \qt\ are similar to \mbox{HD~189733}.

In this work, Section~\ref{Sec:ODR} details the observations and data
reduction processes, Section~\ref{Sec:Analysis} describes the analysis
carried out over the transit light curves, Section~\ref{Sec:TSQ} shows
our results on the transmission spectrum of \qtb, and
Section~\ref{Sec:Concl} gives our discussions and conclusions. We
close this work with Appendix~\ref{Sec:Activity}, containing
information about the activity of \qt.

\section{Observations and data reduction}
\label{Sec:ODR}

\subsection{Observing log and instrumental setup}

\begin{figure}[ht!]
  \centering
  \fbox{\includegraphics[width=.44\textwidth, angle=180]{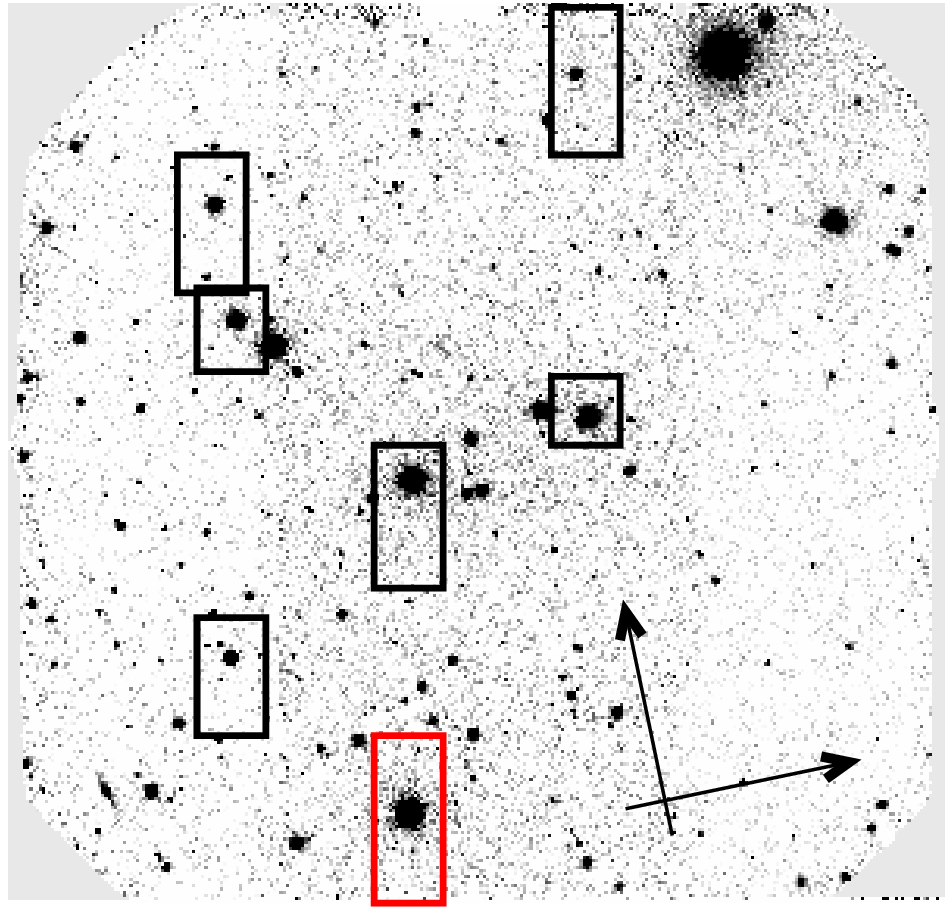}}
  \includegraphics[width=.5\textwidth]{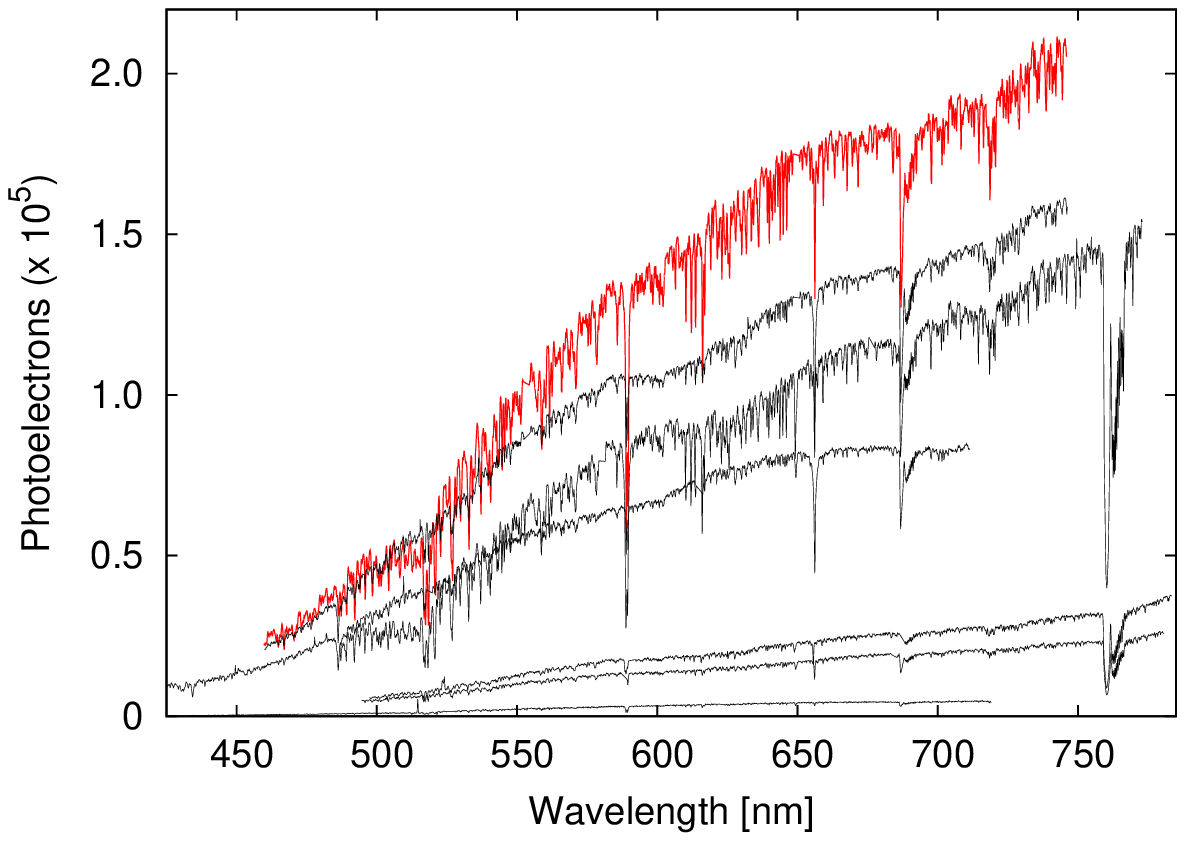}
  \caption{\label{fig:FoVandFlux} {\it Top:} Field of view of our
    observations. Boxes indicate the approximate positions of the
    slits of our costumed mask. North points upward while east
    rightward. Both directions are indicated with arrows. The image
    was acquired using an R filter (r\_G0303). The bright object on
    the very bottom is \qt, which is indicated with a red box. {\it
      Bottom:} Extracted instrumental spectra for \qt\ and six
    reference stars. The different wavelength coverage is caused by
    the uneven distribution of the reference stars at the sky. In both
    figures, Qatar-1 is indicated in red color.}
\end{figure}

\begin{figure}[ht!]
  \centering
  \includegraphics[width=.5\textwidth]{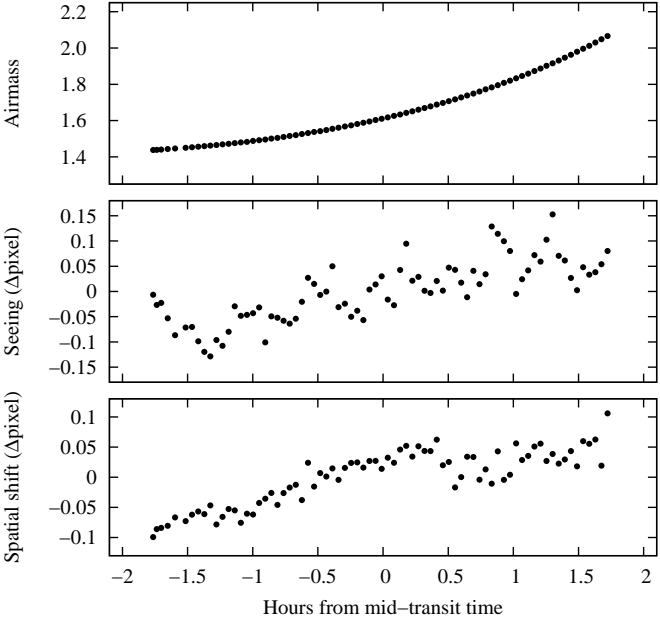}
  \caption{\label{fig:environment} From top to bottom: Air mass,
    seeing and spatial shift in binned pixels during observations of
    \qt\ as a function of the hours from mid-transit time. In the last
    two cases their mean value was subtracted to represent the
    variability as a function of time. The plot sizes were adjusted to
    display the same pixel scale.}
\end{figure}

On the night of September 2, 2014, we observed one transit of
\qtb\ using the 8.1\,m GEMINI North telescope (program Number
GN-2014B-Q-47) and the instrument GMOS in multi-object spectroscopy
mode. The top panel of Figure~\ref{fig:FoVandFlux} shows a pre-image
of the $\sim$7$\times$7 arcmin field of view acquired some days before
the date of the observations. The rectangles indicate the approximate
positions of the slits. From previous photometric follow-ups
\citep{vonEssen2013} we had prior information about the similarity of
the brightness and spectral type between \qt\ and the reference star
inside the central box (\mbox{$\alpha$ = 20:13:41}, \mbox{$\delta$ =
  65:11:33}). Therefore, to maximize our chances of detecting the
planetary atmosphere, the rotation angle of the instrument was chosen
to center both stars in the field. To minimize flux losses we chose to
use wide (15 arcsec) slitlets. The bottom panel of the same Figure
shows the extracted spectra of \qt\ (red), and the six reference stars
(black). Our observations comprise 75 science frames, the last 73
acquired using an exposure time of 150 seconds. Adding to this the
readout time, the overall cadence of our data is 165 seconds. The
science frames were acquired between 08:10:32.7 UT and 12:03:50.0 UT,
adding up to almost four hours of observing time. From the science
frames, 35 were taken during transit and 40 before and after
transit. The air mass ranged between 1.43 (corresponding to a stellar
altitude of 44$^{\circ}$) and 2.06 (29$^{\circ}$). To carry out our
observations we used a \mbox{B600\_G5303} grating centered at
\mbox{589 nm}, providing a wavelength coverage between 460 and
\mbox{746 nm} for centered stars. Figure~\ref{fig:environment} shows
the environmental and observing conditions. Seeing and spatial shift
changes were estimated by fitting a Gaussian profile to a spatial cut
of the flux of \qt\ around the central wavelength. A 4 (spatial)
$\times$ 2 (spectral) binning was used, giving a spatial plate scale
of 0.29 arcsec/pixel; we estimated the mean seeing to be 0.6 arcsec
($\sim 2$ binned pixels).

\subsection{Data reduction}
\label{sec:DR}

The data were reduced using GEMINI/GMOS IRAF tasks in combination with
our own tasks. The steps of the reduction are fully described in
\cite{Trancho2007}, involving overscan and bias subtraction, flat
fielding with Quartz-Halogen flats, and wavelength calibration using a
Cu-Ar lamp spectrum. The latter was acquired using a narrow-slit mask
producing fine well-resolved arc lines. In consequence, the root mean
square of our wavelength solution was well contained within the
resolution of the spectra. Cosmic rays were efficiently cleaned using
the Laplacian Cosmic Ray Identification routine by P. van
Dokkum\footnote{http://www.astro.yale.edu/dokkum/lacosmic/}
\citep{vanDokkum2001}. We performed the background subtraction by
fitting the sky spectrum with a first degree polynomial across the
dispersion using predetermined background regions with the IRAF task
\emph{apall}. Extraction of the one-dimensional spectra was
interactively carried out using the same task, with apertures equal to
1, 2, 3, 5, 8, 10, and 15 times the night average full width at
half maximum (FWHM), determined to be equal to two binned pixels. We
ended up with \mbox{7$\times$7 = 49} one-dimensional spectra for each
time stamp, 7 corresponding to the 7 stars within the field of view,
and 7 corresponding to the previously mentioned apertures.

\section{Model parameters and transit analysis}
\label{Sec:Analysis}

\subsection{Construction of the white light curve} 
\label{sec:wlc_analysis}

The choice of the most suitable reference stars and the respective
aperture was carried out by computing and further analyzing white
light curves integrating the stellar fluxes between 498 and 708
nm. This wavelength range was chosen considering three main aspects:
the flux of all the stars is defined, the signal of all the stars is
relatively large, and atmospheric lines between 720 and 750 nm are
circumvented. Extending the wavelength range would not increase the
photometric precision of the data because toward smaller wavelengths
the signal-to-noise ratio of the spectra decreases significantly
because of the low CCD quantum efficiency, and toward larger
wavelengths there are strong water and ozone absorption lines, which
are expected to change with air mass. Before integrating fluxes, we
checked that there were no wavelength shifts within all the spectra by
analyzing both target and reference stars. To this end, we fitted
Gaussian profiles to the H$\alpha$ line \mbox{($\sim$656 nm)}, the Na
doublet \mbox{($\sim$589 nm)}, and the most prominent Ca line
\mbox{($\sim$621 nm)}. As a reference frame we used the one with the
lowest air mass, which is coincidentally being the first
frame. Averaged differences considering the values obtained from the
center of these three lines between this reference frame and the
remaining frames do not exceed \mbox{$\pm$0.03 nm}. The computed
maximum shift is well contained within one pixel because the natural
resolution of the spectra is about \mbox{0.1 nm}. Since we integrate
in wavelength bins significantly larger than this, we leave the
spectra unshifted. We then produced one light curve per aperture,
dividing the flux of \qt\ to the unweighted averaged fluxes of the
reference stars combined in all possible ways. To each differential
light curve we fitted a \cite{MandelAgol2002} re-binned transit model
\citep[i.e., a transit model computed at a higher cadence but
  re-binned to our time stamps as introduced by][]{Kipping2010} with a
quadratic limb-darkening law, simultaneously with a detrending model
component that accounts for systematics over the light curve (i.e., a
linear combination of time-dependent quantities such as air mass,
pixel shifts, and seeing; for more details see
Section~\ref{sec:detrend}). To minimize computing time, the choice of
reference stars and aperture was carried out minimizing the standard
deviation of the residual light curves which, in turn, were obtained
fitting \cite{MandelAgol2002} transit model times by detrending models
to the data with a least-squares fit. The best reference light curve
was obtained by combining stars 3, 4 and 5, counting from bottom to
top (Figure~\ref{fig:FoVandFlux}), and the corresponding best aperture
was estimated to be \mbox{2$\times$FWHM}.

\subsection{Determination of spectrophotometric errors}
\label{sec:errs}

Once the most suitable reference stars were selected we computed the
individual spectrophotometric errors using the formalism provided by
IRAF's photometric errors, i.e.,

\begin{equation}
  \begin{split}
    \epsilon_i^2 = \frac{\epsilon_{i,\text{Q1}}^2 + \epsilon_{i,\text{RS}}^2}{2}\,\\
    \epsilon_{i,Q1}^2 = (F_{\text{Q1}}/g + A_{\text{Q1}}\sigma_{\text{Q1}}^2 + A_{\text{Q1}}^2\sigma_{\text{Q1}}^2/N_{\text{Q1}})/F_{\text{Q1}}\,\\
    \epsilon_{i,RS}^2 = (F_{\text{RS}}/g + A_{\text{RS}}\sigma_{\text{RS}}^2 + A_{\text{RS}}^2\sigma_{\text{RS}}^2/N_{\text{RS}})/F_{\text{RS}}\,
  \end{split}
\end{equation}

\noindent where Q1 and RS are the errors for \qt\ and the reference
stars, respectively. The parameter $F$ is the integrated flux inside
the wavelength band and aperture, $A$ the area inside the chosen band
and aperture, $\sigma$ the standard deviation of the sky region, $N$
the number of sky pixels, and $g$ the gain of the detector. The
subindex $i$ corresponds to each one of the 73 observations. It is
known that errors produced in this fashion follow a photon-noise-only
distribution. As a consequence, the photometric errors are
underestimated \citep{Gopal-Krishna1995}. Therefore, to produce more
realistic errors we scaled them up to meet the standard deviation of
the residual light curve previously computed. The calculation of
spectrophotometric errors in this way has two advantages: their
averaged magnitudes reflect the scatter of the data and they follow
the air mass trend, increasing in magnitude when air mass increases as
well.

\subsection{Choice of detrending model}
\label{sec:detrend}

As shown in Figure~\ref{fig:environment}, for each time stamp we
measured the air mass, the seeing, and the spatial shifts. Thus, we
considered these quantities while building up the detrending model. To
quantify how much these values correlate with the data, we made use of
the Pearson correlation coefficient, $r_{xy}$. We began subtracting a
transit model to the white light curve, fixing the parameters to those
computed by \cite{Mislis2015}. Then, we calculated the $r_{xy}$
between these residuals and the air mass, giving \mbox{$r_{xy}$ =
  0.96}. A similar exercise was carried out between the residuals and
the seeing and the spatial shift, where in both cases \mbox{$r_{xy}$ =
  0.68}. However, both quantities also vary with air mass, as a result
influencing the $r_{xy}$. Once the air mass trend was subtracted from
these two, the new computed $r_{xy}$ were of 0.11 for the seeing and
0.05 for the spatial shift, corresponding to an almost null
correlation.

Furthermore, we observed a time-dependent sinusoidal modulation, which
can easily be seen by visual inspection of the raw white light
curve. We believe this variability is not related to what
\cite{Stevenson2014} nor \cite{Lendl2016} found, in which case the
systematics correlated with the Cassegrain rotator position angle
(CRPA) and the parallactic angle (PA) of the stars. As performed by
the authors, we extracted the CRPA from the header of our images,
computed the PAs, and found that the frequency at which the cosinus of
these quantities changes is too low (of about a factor of 4) to
represent the data. This variability also does not agree with what
\cite{Sing2012} identified as slit losses, because it does not
correlate with seeing. We further investigated this with GEMINI/GMOS
staff, and although they are aware of this effect, they do not fully
understand its origin (private communication, GEMINI
Helpdesk). Therefore, since we cannot identify the source of our
systematics, to define a proper detrending model and identify which
combination of parameters provides the best representation of the data
we made use of the Bayesian information criterion, \mbox{BIC =
  $\chi^2$ + $k$ ln($N$)}, which penalizes the number $k$ of model
parameters (MP) given \mbox{$N$ = 73} data points, the reduced
chi-squared statistic, $\chi^2_{\mathrm{red}}$, and the standard
deviation of the residual light curves,
$\sigma_{\mathrm{res}}$. Taking into consideration the environmental
and instrumental variability we can measure, plus the sinusoidal
time-dependent variability observed by visually inspecting the white
light curve, as detrending model we considered the following eight
cases:

\begin{enumerate}
\item A linear combination of air mas (AM).
\begin{equation}
   f_1(t) = a_0 + a_1 \text{\it AM}(t)\,. 
   \label{eq:detrending1}
\end{equation}

\item A linear combination between AM and spatial shift (SS).
\begin{equation}
   f_2(t) = a_0 + a_1 \text{\it AM}(t) + a_2 \text{\it SS}(t)\,. 
\end{equation}

\item A linear combination between AM and seeing (SG).
\begin{equation}
   f_3(t) = a_0 + a_1 \text{\it AM}(t) + a_2 \text{\it SG}(t)\,. 
\end{equation}

\item A linear combination between AM, SS and SG.
\begin{equation}
   f_4(t) = a_0 + a_1 \text{\it AM}(t) + a_2 \text{\it SG}(t) + a_3 \text{\it SS}(t)\,. 
\end{equation}

\item A linear combination AM, plus one time-dependent sinusoidal
  function (SIN) with an amplitude $A$, a frequency $\nu$, and a phase
  $\phi$.
  \begin{equation}
  f_5(t) = a_0 + a_1 \text{\it AM}(t) + A_1 \sin[2\piup(t\nu_1 + \phi_1)].\,
  \label{eq:detrending5}
  \end{equation}

\item A linear combination between AM and SS, plus one SIN.
\begin{equation}
  \begin{split}
   f_6(t) = a_0 + a_1 \text{\it AM}(t) + a_2 \text{\it SS}(t) + \\
   A_1 \sin[2\piup(t\nu_1 + \phi_1)]\,.   
  \end{split}
\end{equation}

\item A linear combination between AM and SG, plus one SIN.
\begin{equation}
  \begin{split}
   f_7(t) = a_0 + a_1 \text{\it AM}(t) + a_2 \text{\it SG}(t) + \\
   A_1 \sin[2\piup(t\nu_1 + \phi_1)]\,.   
  \label{eq:detrending7}
  \end{split}
\end{equation}

\item A linear combination between AM, SS and SG, plus one SIN.
\begin{equation}
  \begin{split}
   f_8(t) = a_0 + a_1 \text{\it AM}(t) + a_2 \text{\it SS}(t) + a_3 \text{\it SG}(t) + \\
   A_1 \sin[2\piup(t\nu_1 + \phi_1)]\,.   
  \end{split}
\end{equation}

\end{enumerate}

\noindent Of course, the combination of detrending components can be
as large as desired. However, a larger sample than the one provided
here would be computationally intensive.
Table~\ref{tab:detrend_stats} shows the previously mentioned
statistics that were computed using each one of the detrending
functions, fitted to the data with a simple least-squares minimization
algorithm simultaneously along with a re-binned \cite{MandelAgol2002}
transit model.

\begin{table}[ht!]
  \caption{\label{tab:detrend_stats} Statistics (BIC,
    $\chi^2_{\mathrm{red}}$ and $\sigma_{\mathrm{res}}$) per
    detrending model. The last column denotes the number of model
    parameters, MP. In boldface the two cases considered in this work
    are indicated.}  
  \centering
  \begin{tabular}{l c c c c}
    \hline \hline
    Case &     BIC      &     $\chi^2_{\mathrm{red}}$  &   $\sigma_{\mathrm{res}}$ (ppt) & MP \\
    \hline
    1       & 1350.1       &  19.75             &    0.52                      & 4+2 \\
    2       & 1333.3       &  19.74             &    0.52                      & 4+3 \\
    3       & 1262.5       &  18.63             &    0.51                      & 4+3 \\
    4       & 1248.5       &  18.67             &    0.50                      & 4+4 \\
    {\bf 5} & 209.1        &   2.66             &    0.20                      & 4+5 \\
    6       & 213.7        &   2.71             &    0.20                      & 4+6 \\
    {\bf 7} & 208.6        &   2.65             &    0.20                      & 4+6 \\
    8       & 236.5        &   3.05             &    0.21                      & 4+7 \\ 
    \hline
  \end{tabular}
\end{table}

To begin with, the addition of the sinusoidal variability, between
detrending models 5 to 8 compared to models 1 to 4, provide a
significant reduction in all the statistics. This shows the relevance
of its consideration. Furthermore, focusing on models 5 to 8 we
evidence what we found from their respective $r_{xy}$. Their
statistics show again that the correlation between the data and the
environmental and instrumental quantities is not strong. Overall, the
model does not significantly improve results when compared to the
simplest one of these four, which is just the consideration of air
mass. Since the consideration of detrending function 5 (air mass) and
7 (air mass and spatial shift) provide the best representation of the
data from a statistical point of view, from now on these two cases are
always considered. This also gives us the chance to investigate to
which extent the choice of a detrending model influences our
results. Although the statistics were minimized using $f_7(t)$, the
variability of the spatial shift is well contained within the fraction
of a pixel and thus should not impact our results; the same goes for
the variability of seeing, which is well contained within the fraction
of an arcsecond.

\subsection{Transit fitting}
\label{sec:TFit}

Once the white light curve was fully constructed (time stamps converted
from Julian dates to Barycentric Julian Dates, BJD$_{\rm TDB}$, using
the tools made available by \cite{Eastman2010}, flux and
spectrophotometric errors), and the detrending model components were
determined (functions 5 and 7), we carried out a Markov-Chain Monte
Carlo (MCMC) fitting approach to determine the expectation values of
the orbital and physical parameters of \qtb. In this work all our MCMC
calculations make use of
\texttt{PyAstronomy}\footnote{\url{http://www.hs.uni-hamburg.de/DE/Ins/Per/Czesla/}\\ \url{PyA/PyA/index.html}},
a collection of Python routines providing fitting and sampling
algorithms implemented in the PyMC \citep{Patil2010} and SciPy
\citep{Jones2001} packages.

For the transit we used a \cite{MandelAgol2002} transit model with
quadratic limb-darkening law,

\begin{equation}
  \frac{I(\mu)}{I(1)} = 1 - u_1(1 - \mu) - u_2(1 - \mu)^2,\
\end{equation}

\noindent re-binned to meet the cadence of our data \cite[for
  motivation on the matter, see][]{Kipping2010}. For our custom-model
wavelength-dependent light curves we produced our own limb-darkening
coefficients, $u_1$ and $u_2$. To properly account for any
wavelength-dependent variability introduced by all the optical
components between the source and the CCD, which could in turn affect
the derived limb-darkening values, rather than considering the
laboratory quantum efficiency of the CCD and the transmission of the
grating we used the envelope shape of the observed spectra
directly. Since this shape also includes the continuum emission of
\qt\, we first subtracted a blackbody with parameters matching those
of \qt. To account for the intensity variation of the stellar source
we used angle-resolved PHOENIX spectra \citep{Peter1,Peter2} for a
star of \mbox{T$_{\mathrm{eff}}$ = 4900 K}, \mbox{log g = 4.5} and
\mbox{[Fe/H] = 0.0}, best matching the stellar parameters of
\qt\ \citep{Covino2013}. The PHOENIX spectra were downloaded from the
PHOENIX library\footnote{phoenix.astro.physik.uni-goettingen.de}
\citep{Husser2013}. The limb-darkening coefficients were obtained
fitting the quadratic law to the PHOENIX intensities, previously
convolved with the shape of the observed spectra, and a box function
defined as unity between the minimum and maximum considered wavelength
values, and zero elsewhere. To determine the limb-darkening
coefficients during the fitting procedure we neglected the data points
between \mbox{$\mu$ = 0} and \mbox{$\mu$ = 0.1}, as performed by
\cite{Claret2004}. The derived limb-darkening values can be found
under the third and fourth column of Table~\ref{tab:PARAMS}. A word of
caution: limb-darkening values are given throughout this work with
four decimals of precision. This is obtained after fitting the stellar
intensities with a limb-darkening law. Therefore, this does not
include errors contributed by PHOENIX spectra (errors are not
computed, and therefore the impact is unknown), nor by the intrinsic
errors in the stellar parameters (impact estimated to be on the third
decimal). Limb-darkening values also strongly depend on the specific
intensity spectra used \citep{Csizmadia2013}. Therefore, we caution
against considering any values after the second decimal.

Beside the quadratic limb-darkening law we also computed non-linear
limb-darkening coefficients but found no significant difference in the
residual light curve after the transit fitting was performed. We also
fitted the linear limb-darkening coefficient of the quadratic law to
the data, fixing the quadratic coefficient to the value computed from
PHOENIX intensities. Also, we fitted both limb darkening
coefficients. In all cases we found no significant difference in the
residuals. Since all four scenarios gave fully consistent results, we
finally chose to use the simplest approach. In addition, as
\cite{Mueller2013} pointed out, in the case of a nearly grazing
transit the planet does not cross the center of the star. Therefore,
the transit light curve does not contain sufficient information
concerning the limb-to-center brightness variation to fit for the
limb-darkening coefficients. Hence, throughout this work we fix the
coefficients to theoretical values. For the majority of exoplanet host
stars the empirical limb-darkening coefficients derived from transit
light curves match their theoretical counterparts reasonably well
\citep{Mueller2013}. However, this approach might introduce
systematics in the transmission spectrum, as shown already by the
significant discrepancy between fitted and theoretical limb-darkening
coefficients observed by \cite{Claret2009} (\mbox{HD~209458}) and
\cite{Mallonn2016} (\mbox{HAT-P-32}).

For the transit model the fitting parameters are the semi-major axis
scaled by the stellar radius, $a/R_\mathrm{S}$, the orbital
inclination measured with respect to the plane of the sky, i, the
mid-transit time, T$_\mathrm{0}$, and the planet-to-star radii ratio,
R$_\mathrm{P}$/R$_\mathrm{S}$. The first three parameters are
wavelength independent. The quadratic limb-darkening coefficients,
$u_1$ and $u_2$, were considered fixed, and are considered fixed
throughout this work. Since we count with one transit, the orbital
period was considered as fixed to the value determined by
\cite{Mislis2015}. Simultaneously to the transit model we implemented
the two detrending models given by $f_5(t)$ and $f_7(t)$, which are
fully described in Eq.~\ref{eq:detrending5} and
Eq.~\ref{eq:detrending7}.

The fitting procedure was carried out in two stages. As starting
values for the fitting parameters we used those listed in
\citep{Mislis2015} for the transit model, and our custom
limb-darkening coefficients. We chose conservative uniform priors for
all the parameters. First, we iterated 1.5 $\times$10$^6$ and burned
the initial 50\ 000 samples. Then, analyzing the posterior
distributions, from their mean and standard deviations we found the
expectation values of the orbital parameters and corresponding errors,
respectively. Thus, throughout this work, errors on the parameters are
at 1$\sigma$ level. Using these values we computed a best-fit model
and the residuals by simply subtracting the data to the best-fit
model.

\subsection{Correlated noise treatment}

To quantify to which extent the residual light curve is affected by
correlated noise, we followed a similar approach as described in
\cite{Gillon2006,Winn2008}, and \cite{Carter2009}. We started dividing
the residual light curve into $M$ bins of 15 minutes each, which
corresponds to the approximate duration of ingress/egress. Then, we
calculated the mean value of data points per bin, N. If the data are
affected by correlated noise, the sample standard deviation of the
binned data, $\sigma_N$, differ by a factor \mbox{$\beta_N$} from its
theoretical expectation \citep[see, e.g.,][for a more extended
  description]{vonEssen2013}. For data sets free of correlated noise,
\mbox{$\beta_N$ = 1} is expected. The consideration of $f_5(t)$ and
$f_7(t)$ resulted in no measurable correlated noise. If $\beta_N$ had
been different from 1, we would continue by enlarging the
spectrophotometric errors by $\beta_N$, and by carrying out the
transit fitting procedure all over again. For the white light curve
the amount of correlated noise was negligible and, in consequence,
this step was not necessary.

\subsection{Transit parameters from the white light curve}

\begin{figure}[ht!]
  \centering
  \includegraphics[width=0.5\textwidth]{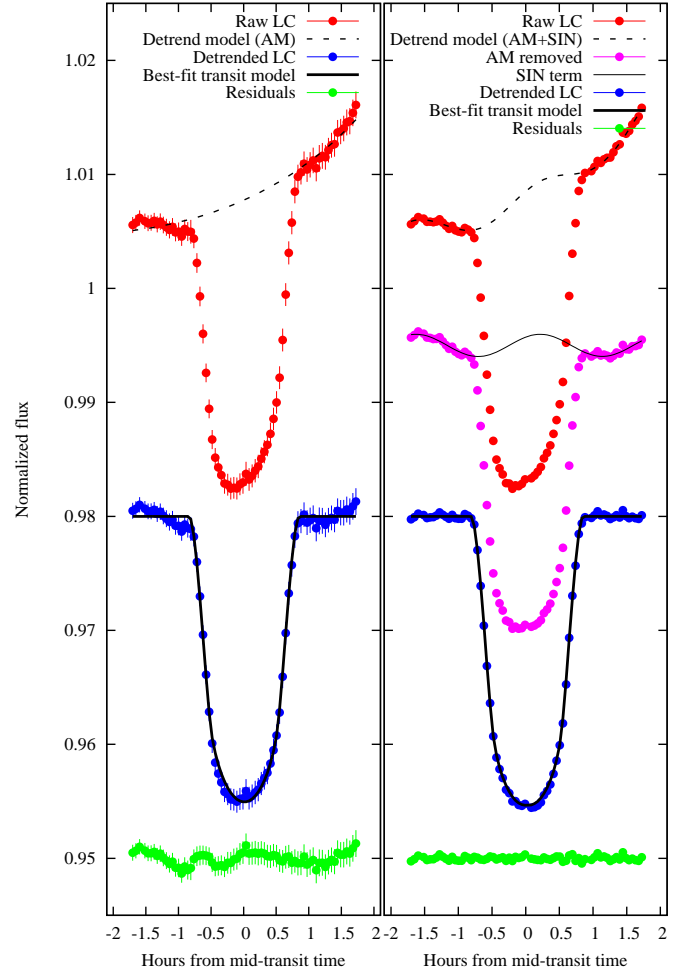}
  \caption{\label{fig:WLCs} White light curve of \qtb. From top to
    bottom, the left panel of the Figure shows the raw light curve in
    red. The air-mass model is overplotted in dashed
    lines. Artificially shifted below, the white light curve after
    subtracting the air-mass model is indicated in blue and the
    best-fit transit model is shown in a black continuous
    line. Finally, the residuals are shown in green. Error bars in all
    figures were enlarged to meet the standard deviation of the
    residuals. The right panel of the Figure shows the raw white light
    curve in red points, along with the best-fit transit and
    detrending models in black dashed line for $f_5(t)$. The pink
    light curve corresponds to the raw data when the best-fit air-mass
    component has been subtracted, along with the sinusoidal
    time-dependent variability overplotted in a black thin continuous
    line. Blue points correspond to a fully detrended transit light
    curve; the best-fit transit model is overplotted in a black thick
    line. Below in green, the residual light curve obtained
    subtracting the best-fit transit model to the blue points
    (equivalently, subtracting the dashed line to the red points) can
    be seen. Everything has been artificially shifted.}
\end{figure}

The best-fit transit parameters of \qtb\ are summarized in
Table~\ref{tab:PARAMS} for both detrending models. The white light
curves obtained considering $f_1(t)$ and $f_5(t)$, along with the
model components, are plotted in Figure~\ref{fig:WLCs} to stress the
relevance of adding the sinusoidal term. The standard deviation of the
residual white light curve is of 0.18 parts per thousand (ppt), one of
the most precise ground-based white light curves ever observed.

\subsection{Considerations on limb-darkening treatment}

In order to test if there is an impact on the transit parameters
caused by fixing the limb-darkening coefficients we repeated the
exercise described here but added the linear limb-darkening
coefficient as the fitting parameter, and in a further step both
linear and quadratic coefficients. We found that the best-fit transit
parameters are fully consistent with those shown in
Table~\ref{tab:PARAMS} when errors at 1$\sigma$ level are
computed. This is in good agreement with \cite{Espinoza2015}. In this
work, the authors found no significant difference in the
planet-to-star radii ratio nor in the semi-major axis around 4900
Kelvin (\mbox{Qatar-1's} effective temperature) when limb-darkening
coefficients were either fixed or fitted (see their Figure 10). A
comparison of the transit parameters derived in this work to values
published in the literature is given in Table~\ref{tab:PARAMS}. This
comparison shows an unusual large discrepancy of the transit
parameters among different studies. This might be caused by the
different treatment of the limb-darkening coefficients, as, for
example, the \cite{Covino2013} fits for the linear limb-darkening
coefficients while they are fixed to theoretical values in this
work. As mentioned before, the fitting of the limb-darkening
coefficients for nearly-grazing planets leads to erroneous
parameters. Differences in the calculation of theoretical values can
cause systematic differences in the limb-darkening coefficients, too,
translating into differences in the transit parameters. A homogeneous
study of all transit data is beyond the scope of this work.

\begin{table*}[ht!]
  \caption{\label{tab:PARAMS} Our best-fit orbital parameters for the
    white light curve, along with 1$\sigma$ errors compared to our
    previous work. The orbital period was adopted from
    \cite{Mislis2015}. The mid-transit time, T$_\mathrm{0}$, is in
    BJD$_{\mathrm{TDB}}$-2456000.}  \centering \scalebox{0.9}{
  \begin{tabular}{l c c c c c}
    \hline \hline
    Parameter    & This work, $f_5(t)$, $f_7(t)$   &   \cite{Alsubai2011}       &   \cite{vonEssen2013}      &  \cite{Covino2013}  &  \cite{Mislis2015}  \\    
    \hline
    a/R$_S$       & 6.59 $\pm$ 0.04                &    6.101 $\pm$ 0.067       &  6.42 $\pm$ 0.10          &  6.24 $\pm$ 0.09     &  6.25 $\pm$ 0.08 \\    
    i ($^{\circ}$) & 84.48 $\pm$ 0.10               &    83.47$^{+0.40}_{-0.36}$     &  84.52 $\pm$ 0.24         &  83.82 $\pm$ 0.25    &  84.03 $\pm$ 0.16   \\
    R$_P$/R$_S$   & 0.1523 $\pm$ 0.0004            &    0.1455 $\pm$ 0.0015      &  0.1435 $\pm$ 0.0008     &  0.1513 $\pm$ 0.0008  &  0.1475 $\pm$ 0.0009 \\ 
    T$_\mathrm{0}$ & 902.93413 $\pm$ 0.00006       &   & & & \\
    $\sigma_{\mathrm{res}}$ (ppt) & 0.18             &   & & & \\
    u$_1$         &   0.6216                      &   & & & \\
    u$_2$         &   0.1119                      &   & & & \\
    \hline
  \end{tabular}
  }
\end{table*}

\begin{table*}[ht!]
  \caption{\label{tab:results} Values derived analyzing chromatic
    light curves while considering the detrending model $f_5(t)$. From
    left to right; WC corresponds to the wavelength channel (in nm),
    indicating the beginning and end of the wavelength band; CW
    corresponds to the channel width in nm; u$_1$, and $u_2$ show the
    custom limb-darkening coefficients; R$_\mathrm{P}$/R$_\mathrm{S}$
    the best-fit transit depth along with their derived 1$\sigma$
    errors; $a_0$, $a_1$, $A_1$, $\nu_1$ and $\phi_1$ correspond to
    the detrending coefficients as specified in
    Eq.~\ref{eq:detrending5}; SDR corresponds to the standard
    deviation of the residual light curve in ppt; and $\beta_N$
    accounts for correlated noise in the light curves. For the
    detrending parameters we specify their precision by adding a
    parenthesis in their last significant decimal.}
  \centering
  \begin{tabular}{l c c c c c c c c c c c}
    \hline \hline
    WC  & CW  &  $u_1$   &   $u_2$   & R$_\mathrm{P}$/R$_\mathrm{S}$ & $a_0$ & $a_1$ & $A_1$ & $\nu_1$ & $\phi_1$ & SDR & $\beta_N$\\
    \hline
5100 - 5500 & 20.0 & 0.7527 & 0.0385 & 0.1529 $\pm$ 0.0003 & 1.008(9) & 0.014(5) & 0.001(2) & 12.(4) & 0.5(3) & 0.27 & 1.00 \\
5500 - 5750 & 12.5 & 0.6922 & 0.0883 & 0.1521 $\pm$ 0.0006 & 1.008(6) & 0.016(7) & 0.000(9) & 13.(2) & 0.7(1) & 0.46 & 1.00 \\
5750 - 6000 & 12.5 & 0.6579 & 0.1049 & 0.1527 $\pm$ 0.0003 & 1.008(6) & 0.016(7) & 0.000(9) & 12.(9) & 0.5(3) & 0.26 & 1.00 \\
6000 - 6100 &  5.0 & 0.6337 & 0.1143 & 0.1529 $\pm$ 0.0004 & 1.008(6) & 0.016(7) & 0.001(0) & 13.(0) & 0.5(0) & 0.42 & 1.15 \\
6100 - 6250 &  7.5 & 0.6106 & 0.1203 & 0.1533 $\pm$ 0.0005 & 1.008(6) & 0.016(7) & 0.001(1) & 12.(0) & 0.3(6) & 0.38 & 1.04 \\
6250 - 6385 &  6.7 & 0.6013 & 0.1228 & 0.1522 $\pm$ 0.0004 & 1.008(6) & 0.016(7) & 0.001(0) & 13.(1) & 0.8(7) & 0.29 & 1.30 \\
6385 - 6500 &  5.7 & 0.5862 & 0.1262 & 0.1519 $\pm$ 0.0004 & 1.008(6) & 0.016(7) & 0.001(0) & 13.(1) & 0.8(7) & 0.46 & 1.15 \\
6500 - 6670 &  8.5 & 0.5582 & 0.1441 & 0.1526 $\pm$ 0.0003 & 1.008(6) & 0.016(7) & 0.001(0) & 13.(4) & 0.6(4) & 0.25 & 1.00 \\
6670 - 6800 &  6.5 & 0.5527 & 0.1399 & 0.1519 $\pm$ 0.0005 & 1.008(6) & 0.016(7) & 0.000(9) & 12.(9) & 0.5(3) & 0.38 & 1.00 \\
6800 - 6930 &  6.5 & 0.5408 & 0.1437 & 0.1525 $\pm$ 0.0006 & 1.008(6) & 0.016(7) & 0.000(9) & 11.(9) & 0.9(1) & 0.32 & 1.48 \\
6930 - 7000 &  3.5 & 0.5349 & 0.1439 & 0.1514 $\pm$ 0.0003 & 1.008(6) & 0.016(7) & 0.000(9) & 13.(0) & 0.9(0) & 0.36 & 1.00 \\
7000 - 7080 &  4.0 & 0.5238 & 0.1448 & 0.1514 $\pm$ 0.0004 & 1.008(6) & 0.016(7) & 0.000(9) & 12.(9) & 0.5(7) & 0.36 & 1.00 \\
\hline
  \end{tabular}
\end{table*}

\section{Analysis and results}
\label{Sec:TSQ}

\subsection{Construction of the wavelength-dependent light curves}

Rather than dividing the full wavelength range in equally sized bins,
we estimated the amount and extension of the wavelength channels
following a quality criteria. Assuming that the probed atmosphere of
the exoplanet has a maximum variability of 3 scale heights within our
wavelength range \citep{Fortney2010}, considering a nominal value of
\mbox{R$_\mathrm{P}$/R$_\mathrm{S}$ = 0.1522}, and an average
temperature of 1400 Kelvin \citep{Covino2013}, this would be
translated into a maximum variability of R$_\mathrm{P}$/R$_\mathrm{S}$
of about 1.4 ppt. If our goal is to achieve a 2$\sigma$ detection, the
error bars on \mbox{R$_\mathrm{P}$/R$_\mathrm{S}$} should be as large
as 0.5 ppt. To choose the number of wavelength channels taking this
into consideration we carried out the following exercise: using
exactly the same time stamps as our measurements, and using as the
best-fit values of the white light curve and the $f_5(t)$ detrending
model as input parameters, we simulated light curves with a given
amount of white noise. The first standard deviation considered in this
exercise was that given by the standard deviation of the white light
curve, \mbox{0.18 ppt}, and 2, 4, 6, 8, and 10 times this value. To
make this exercise as realistic as possible, rather than considering
constant values for the error bars, we used the real errors of the
white light curve but enlarged these errors to meet the noise level of
the synthetic data. Then, following the approach described in
Section~\ref{sec:TFit}, we fitted the orbital parameters
a/R$_\mathrm{S}$, i, T$_\mathrm{0}$ and
R$_\mathrm{P}$/R$_\mathrm{S}$. To accomplish this, we used Gaussian
priors for the first three parameters, since we know them quite
precisely from the white light curve analysis and they are wavelength
independent, and a uniform prior distribution for
R$_\mathrm{P}$/R$_\mathrm{S}$. After 10$^5$ iterations and a discard
of the first 25\%, we computed as usual 1$\sigma$ errors on
R$_\mathrm{P}$/R$_\mathrm{S}$. Our simulations show that the standard
deviation that would limit our required precision is about \mbox{0.5
  ppt}. Since these light curves only have Gaussian noise, we consider
the derived precision as an upper limit. In consequence, we defined
the bin number and wavelength region so that the standard deviation of
each light curve is \mbox{$\sim$ 0.5 ppt} or lower, which leads to the
12 wavelength bins presented in this work. While following this
approach we checked that the choice of wavelength channels and this
integration scheme does not make a wavelength bin fall right in the
middle of a Fraunhofer line, which might cause variability extrinsic
to the planetary atmosphere. The first column of
Table~\ref{tab:results} shows the wavelength range for each one of the
12 wavelength bins, while the second column of the same Table
indicates the bin size, both in nm. The total wavelength range was
reduced forward, starting from 510 nm and ending in 708 nm, rather
than the original 498-708 nm range. This was chosen to minimize
correlated noise in the bluest light curve.

\subsection{Analysis of the chromatic light curves}

Once the wavelength bins were defined, to derive the transmission
spectrum of \qtb\ we carried out a similar approach to the white light
curve analysis regarding the calculation of the spectrophotometric
errors, the $\beta_N$ factors, limb-darkening coefficients, and
detrending functions, and we used the same choice of reference stars
and aperture as determined in Section~\ref{sec:wlc_analysis}. In all
cases, R$_\mathrm{P}$/R$_\mathrm{S}$ were fitted with a uniform
probability density function, while a/R$_\mathrm{S}$, i and
T$_\mathrm{0}$ had Gaussian probability density functions with mean
and standard deviation equal to the best-fit and error values obtained
from the white light curve, respectively. While the
\mbox{R$_\mathrm{P}$/R$_\mathrm{S}$} were fitted to each light curve
individually, a/R$_\mathrm{S}$, i and T$_\mathrm{0}$ were fitted
simultaneously to all the chromatic light curves. In other words,
these three transit parameters best-fit all the light curves
simultaneously. In some works, at this stage the
wavelength-independent parameters are fixed. However, since it is our
intention to compute reliable error bars for
\mbox{R$_\mathrm{P}$/R$_\mathrm{S}$}, we fully used the information
obtained from the white light curve and properly propagated their
errors into the computation of the wavelength-dependent
\mbox{R$_\mathrm{P}$/R$_\mathrm{S}$}'s.

With respect to the detrending coefficients, $a_0$ and $a_1$ for the
air-mass component, and $A_1$, $\nu_1$ and $\phi_1$ for the sinusoidal
component, we tried several fitting procedures. First, we considered
one set of independent detrending parameters per chromatic light
curve. In this case the detrending parameters were \mbox{5$\times$12 =
  60}. Visually inspecting our resulting light curves we found that,
when placing them all together, the air-mass component was the same
for all the chromatic light curves. We believe this is because the
colors of the brightest reference star and \qt\ are similar
\citep{vonEssen2013} and, thus, color differences produced from the
wavelength-dependent absorption of stellar light produced in our
atmosphere are minimized. This similarity was not observed when the
sinusoidal component of the detrending model was inspected. We fitted
one $a_0$ and one $a_1$ simultaneously to all the light curves to
reduce the parameter space in a sensible way and minimize the impact
they might have over the retrieved transmission spectrum. Following
\cite{Gibson2013b} and \cite{Lendl2016}, we also tried to subtract the
common noise to all the light curves, but we found that this is not
sufficient since there appears to be a wavelength dependency with the
sinusoidal component of the model. When visually inspecting the
residual light curves we noted that this approach was insufficient for
the bluest light curve, which is where the atmosphere plays a larger
role. Therefore, for this light curve we fitted a set of ($a_0$,
$a_1$) individually, while we fitted another equivalent set of
parameters to the remaining data simultaneously. As previously
mentioned, all these trial exercises were carried out considering the
two detrending functions $f_5(t)$ and $f_7(t)$, without influencing on
the results. In this Section we limit ourselves to show parameters
derived from $f_5(t)$.

Table~\ref{tab:results} summarizes our results, listing the wavelength
bins and sizes, computed limb-darkening coefficients, derived
$\mathrm{R_P/R_S}$ values along with 1$\sigma$ errors, fitting
parameters, $\beta_N$ values, and standard deviation of the
residuals. Figure~\ref{fig:transits} shows the 12 detrended transit
light curves on the left panel, their respective best-fit transit
model in continuous black line, and the residual light curves for the
different wavelength channels on the right panel. Each light curve has
been artificially shifted to allow for visual inspection.

\begin{figure*}[ht!]
  \centering
  \includegraphics[width=.9\textwidth]{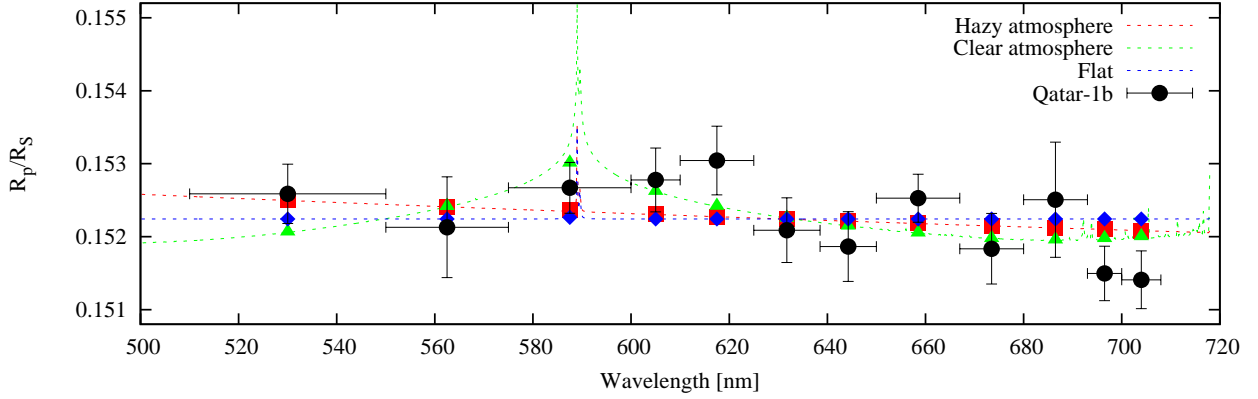}
  \caption{\label{fig:TS} Transmission spectrum of Qatar-1b. Black
    circles and horizontal error bars show the derived
    R$_\mathrm{P}$/R$_\mathrm{S}$'s and their
    uncertainties. Horizontal lines are not error bars but indicate
    the size of the wavelength bin. Green, red and blue continuous
    lines correspond to \cite{Fortney2010} models for exo-atmospheres
    investigated in this work. Filled diamonds, squares and triangles
    indicate averages of the models within each wavelength bin. The
    models were artificially shifted to meet the data.}
\end{figure*}

\begin{figure*}[ht!]
  \centering
  \includegraphics[width=.9\textwidth]{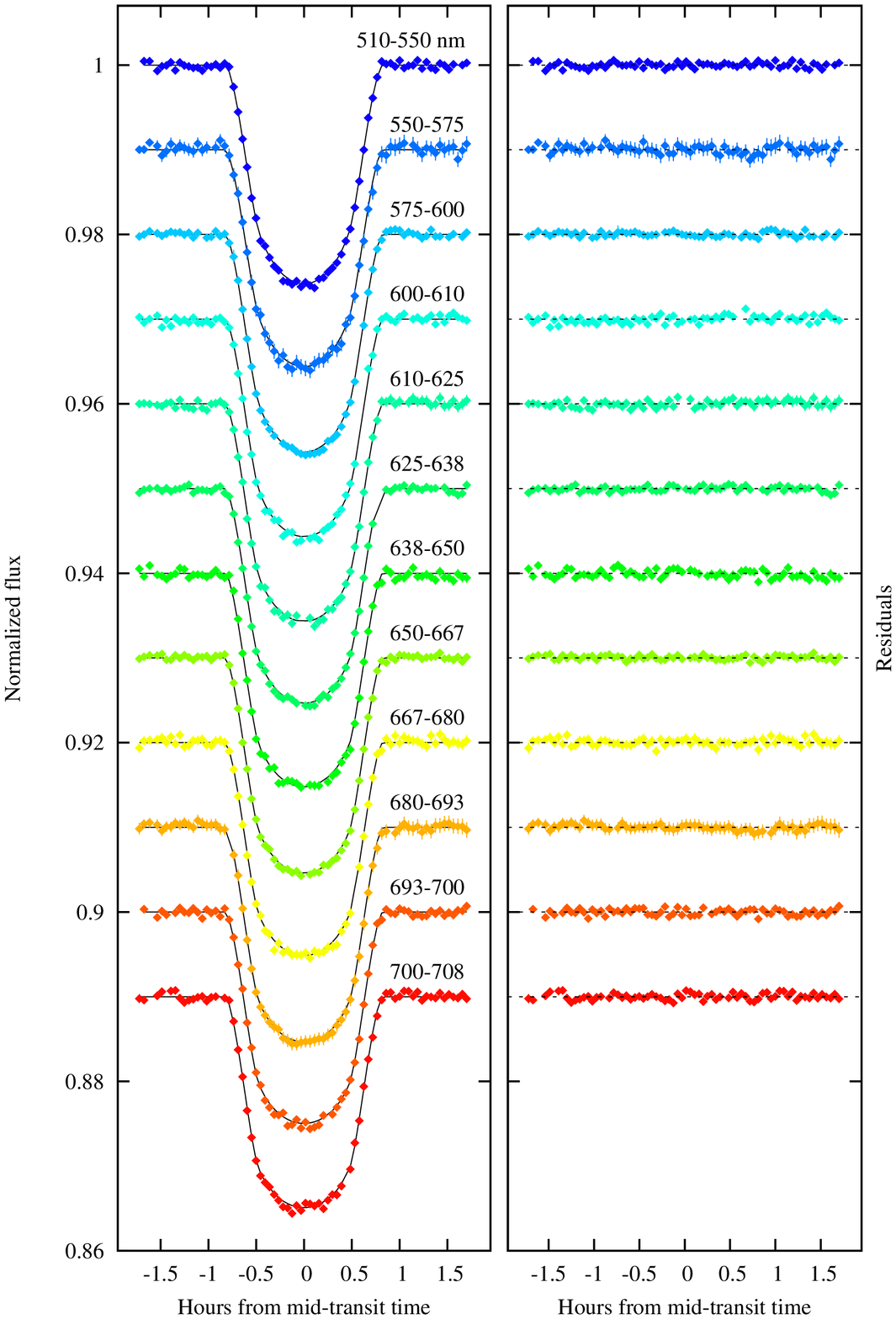}
  \caption{\label{fig:transits} Our 12 chromatic and detrended primary
    transit light curves shown in the left panel along with their
    respective best-fitting transit models in black continuous
    lines. The right panel shows the residual light curves obtained by
    subtracting the best-fit models (transit times detrending) to the
    raw data. All light curves were artificially shifted to fit the
    plot.}
\end{figure*}

\subsection{Transmission spectrum of Qatar-1b}

Figure \ref{fig:TS} shows the transmission spectrum of \qtb\ obtained
following the steps detailed in the previous section. When we used the
detrending function $f_5(t)$, along with theoretical models. $f_7(t)$
gave fully consistent results. We compared our transmission spectrum
to several theoretical models \citep{Fortney2010} that were computed
for an atmospheric temperature of 1500~K and surface gravity of 25
m/s$^2$, and we scaled them to the planetary parameters of \qtb. In
the Figure, continuous lines show the original models. The models were
averaged within the same wavelength channels as the observed
transmission spectrum of \qtb. Regarding the Figure, the only fitted
parameter is a vertical offset necessary to match observations to
models. The first model is a solar-composition model with TiO
artificially removed (green triangles and dashed line). The second
model is a solar composition model without TiO, which has a Rayleigh
scattering component with a cross section a thousand times that of
H$_2$ (red squares and dashed line). The third model includes a gray
absorber that cuts all features above a certain pressure resulting in
a nearly flat spectrum (blue diamonds and dashed line). The fourth
model has as main absorber TiO. Since this last model provides the
poorest representation of the observed transmission spectrum and has
too many features that block the rest, we did not add it into the
Figure. To assess which model best represents the data, we computed
$\chi^2_{\mathrm{red}}$ and its respective $P$ value for each model,
considering 12-1 degrees of freedom. The significance level is chosen
to be 5\%. Both statistics are summarized in Table~\ref{tab:stats}.

\begin{table}[ht!]
  \caption{\label{tab:stats} Computed statistics for the atmospheric
    models tested in this work.}  
  \centering
  \begin{tabular}{l c c}
    \hline\hline
    Model            & $\chi^2_{\mathrm{red}}$ &    $P$    \\
    \hline
    Hazy atmosphere  &    1.161      &  0.308 \\
    Clear atmosphere &    1.063      &  0.386 \\
    Flat             &    1.569      &  0.100 \\
    TiO dominant     &    2.069      &  0.019 \\
    \hline
  \end{tabular}
\end{table}

As in the Figure, the computed $\chi^2_{\mathrm{red}}$, and the $P$
value show that the hazy and clear atmosphere models are in agreement
with our data, and the flat and TiO dominant models are
disfavored. Although this is not common in hot Jupiter atmospheres
\citep{Sing2016}, our statistical (and very simple) analysis favors
the clear atmosphere. Therefore, we caution against reaching any
conclusion about the composition of the atmosphere of \qtb\ from this
data alone. We unfortunately lack the precision necessary to make a
comparative study between the atmospheres of \qtb\ and
\mbox{HD~189733b}. More extensive wavelength coverage and space-based
data are required.

\section{Discussion and conclusions}
\label{Sec:Concl}

In this work we report GEMINI-N/GMOS spectroscopic observations
carried out during one transit of \qtb\ covering the wavelength range
between 460 and 746 nm. Our main goal was to study the transmission
spectrum of \qtb\ and compare it to the well-known
\mbox{HD~189733b}. The observations were motivated by the
compatibilities in the orbital and physical parameters of both
systems. It was our intention to analyze the possibility of exoplanet
atmospheres to be dominated by their environments, which could be
answered with follow-ups in similar systems, such as the two mentioned
here.

The data collected and investigated in this work allowed us to refine
the transit parameters. We derived an optical transmission spectrum of
the planet that included the sodium line by creating 12 chromatic
light curves. The transmission spectrum of \qtb\ was extracted via
wavelength channels whose width varied between 3.5 and 20 nm, which
were chosen to minimize the standard deviation of the individual light
curves and to circumvent the Fraunhofer lines. Owing to the nearly
grazing orbit of \qtb, during our data analysis we did not fit the
limb-darkening coefficients. We customized our own limb-darkening
values by convolving PHOENIX angle-resolved intensities with the
transmission of the full optical setup, to have maximum control over
the limb-darkening related systematics.

The observations took place around BJD$\sim$2456903. From the
ephemeris found by our H$\alpha$ photometric follow-up (see next
Appendix and Figure~\ref{fig:follow_up}) the transit observed here
took place close to a maximum of flux which, in turn, corresponds to a
minimum in spot coverage \citep[this was cross-matched to the spot
  modulation observed by][]{Mislis2015}. \cite{Alsubai2011} did not
find any spot crossing events in their survey photometry, nor did
\cite{Mislis2015} in their high-precision photometric follow-up. The
authors suggested that the planet could be crossing latitudes of the
star showing low spot activity. In an attempt to better characterize
the system, we carried out a photometric follow-up of the host star in
H$\alpha$ using the 1.2~m telescope located in Hamburg, Germany. We
see clear evidence of activity correlated with what we estimated to be
the rotational period of the star. From further spectroscopic
observations we confirmed \qt\ to be a moderately active star, in
agreement with \cite{Covino2013} and \cite{Mislis2015}. In this
context, modifications in the depth of transit light curves due to
spots can occur. To quantify the amplitude of this effect,
\cite{Sing2011b} have already characterized that a decrease of 1\% in
the stellar flux (slightly larger than the H$\alpha$ variability
reported here) would increase the transit depth by about 1\% (their
Figure 10). This would be translated into a maximum uncertainty in the
radius ratio of 0.0002. However, as calculated by, for example,
\cite{Mallonn2015} for the similar host star HAT-P-19, the chromatic
effect on the transmission spectrum is about an order of magnitude
lower. We conclude that unocculted star spots do not significantly
modify the transmission spectrum derived in this work.

After a careful analysis of our chromatic light curves, we find that
the wavelength coverage and the precision of the transmission spectrum
is not sufficient to either rule out or strongly favor classic
atmospheric models. The simple statistical analysis carried out in
this work seems to favor the clear atmosphere scenario. However, we
caution against reaching any conclusions from our data alone. A larger
wavelength coverage and space-based data would be required to
characterize the constituents of the atmosphere of \qtb.

\appendix

\section{Stellar activity}
\label{Sec:Activity}

\subsection{H$\alpha$ photometric follow-up}

The H$\alpha$ chromospheric emission is known to be one of the primary
indicators of activity and magnetic heating in low-mass stars
\citep[see, e.g.,][for a characterization of activity indicators for
  spectral types from F to M]{Cincunegui2007}. Therefore, to
characterize the activity level of \qt, we followed up
\qt\ photometrically during two months using the 1.2~m Oskar L\"uhning
Telescope (OLT) located at Hamburger Sternwarte in H$\alpha$ light.

\begin{figure}[ht!]
  \centering
  \includegraphics[width=0.5\textwidth]{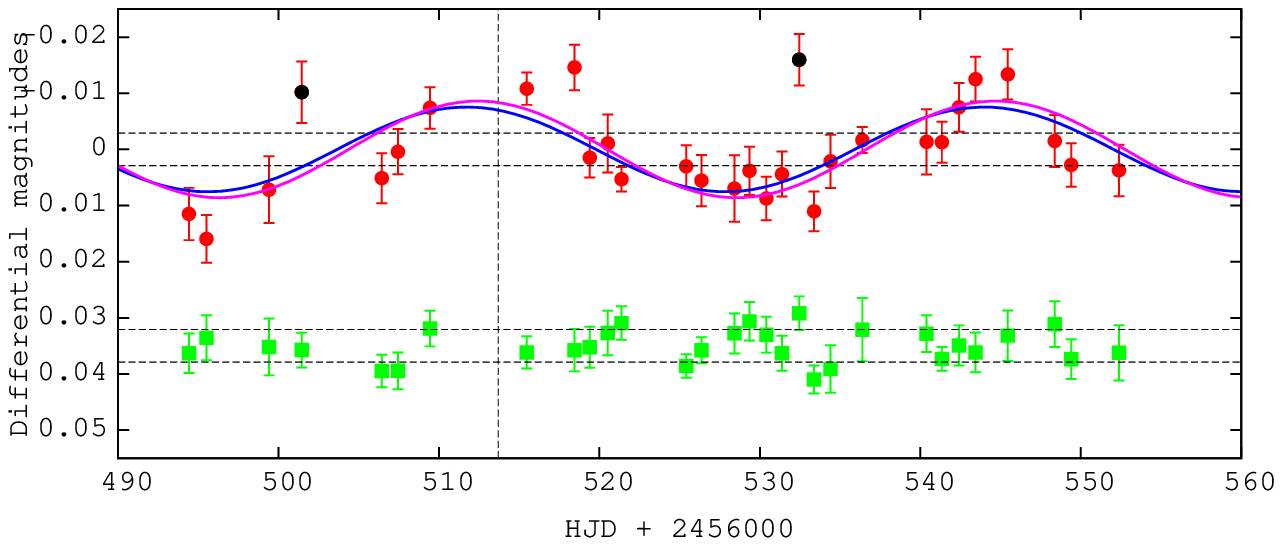}
  \caption{\label{fig:follow_up}H$\alpha$ photometric follow-up of
    \qt. The differential light curve for \qt\ is indicated on the top
    in red circles, while the control light curve is indicated on the
    bottom in green squares. Horizontal lines show $\pm 1\sigma$ the
    scatter of the control light curve. The blue and pink lines show
    the best-fit sinusoidal to the data. The vertical dashed line
    shows the position at which we also acquired spectral information
    on Qatar-1.}
\end{figure}

For each one of the observing nights obtained with the OLT, we
calculated the mean and the dispersion of the data and considered
these as representative values for the flux and the intra-night
scatter. These values are shown in Figure~\ref{fig:follow_up}, where
differential magnitudes are plotted as a function of time. On the
bottom of the figure, plotted in green is the control differential
light curve, which has been artificially shifted. Plotted in red
circles, on top, is the differential light curve of \qt. The scatter
of the two overall light curves satisfies
\mbox{$\sigma_{\mathrm{Qatar-1}}\sim 3 \sigma_{\mathrm{Control}}$},
indicating that the variability of \qt\ may be inherent. The
horizontal dashed lines in Fig.~\ref{fig:follow_up} indicates a band
of 2$\sigma_{\mathrm{Control}}$ width around the mean.

We noticed two points, indicated as black dots, which appear to be
outside the data distribution. One can address this increase in the
flux due to activity on the surface of the star or a color-dependent
atmospheric effect. However, we can not ensure the truthfulness of the
data points beyond the stability of the control light curve.

We applied a Lomb-Scargle periodogram \citep{lomb,Scargle,LombScargle}
to search for any periodicity contained in \qt\ light curve and to the
control light curve for a sanity check. In the first case, we found a
significant peak at \mbox{$\nu_{\mathrm{Q1}}$ = 0.033 $\pm$ 0.005 c/d}
\mbox{($P_{\mathrm{Q1}}$ = 30 $\pm$ 7 days)}. The error for the
frequency corresponds to the dispersion $\sigma_{\mathrm{Gauss}}$ of a
Gaussian function, which was fitted to the leading peak. The
false-alarm probability (FAP) of the maximum power is
0.002\%. \cite{Covino2013} confirmed the star to be a slow rotator
\mbox{($v \sin(i)$ = 1.7 $\pm$ 0.3 km/s)}. Assuming the reported
\mbox{$v \sin(i)$} to be the speed at the stellar equator, and
considering the stellar radius estimated by \cite{Alsubai2011}
\mbox{($\mathrm{R_S}$ = 0.823 $\pm$ 0.025 $R_{\odot}$)}, the observed
velocity would translate into a rotational period of \mbox{$\sim$ 25
  $\pm$ 5 days}. Within errors, both periods seem to be consistent. We
interpret $\nu_{\mathrm{Q1}}$ to be the rotational period of the star,
which is in good agreement with that reported by \cite{Mislis2015}.

Once we determined the periodicity within the data, we fitted a
sinusoidal variation to \qt\ data in the form,

\begin{equation}
  H_{\alpha}(t) = \Delta ~\mathrm{mag} \cdot \sin(2\pi(t\cdot \nu_{Q1} + \phi))\; ,
\end{equation}

\noindent where $\Delta$mag is the magnitude variation, $\phi$ the
phase, and $\nu_{\mathrm{Q1}}$ considered as fixed in the value
already mentioned. After fitting the complete data set first, and the
data set without the points outside the distribution, we found that
changes in the fitted amplitudes are contained within the precision of
the data (blue and pink continuous lines,
Fig.~\ref{fig:follow_up}). Therefore, we estimated the photometric
H$\alpha$ variability as \mbox{$\Delta$mag = 0.009 $\pm$
  0.001}. According to Pogson's law \citep{ALLEN}, this equates to a
flux variation of $\sim$0.8\%.

\subsection{Stellar activity and age}

Adopting our estimated rotational period for \mbox{Qatar-1},
\mbox{$P_{\mathrm{Q1}}$ = 30 $\pm$ 7 days}, we can estimate the
gyrochronological age, $t_{\mathrm{gyro}}$, using the relation given
by \cite{Barnes2007}, Eq. 3. Considering a $T_{\mathrm{eff}}$ of 4910 K
\citep{Covino2013}, the corresponding color index is (B-V) = 0.9. We
obtain \mbox{$t_{\mathrm{gyro}}$ = 2.7 $\pm$ 1 Gyr}.

\subsection{The H$\alpha$ equivalent width as activity indicator}

We observed \qt\ spectroscopically using the Hamburg Robotic
Telescope
(HRT\footnote{\url{http://www.hs.uni-hamburg.de/DE/Ins/HRT/hrt_main.html}})
located at La Luz observatory, in Guanajuato, Mexico
\citep{Mittag2011}. The telescope has a primary mirror of 1.2~m and a
fibre-fed Echelle spectrograph. The spectral distribution is divided
in a blue and a red channel, covering the wavelength range between
$\sim$380 to $\sim$880 nm. The spectral resolution is estimated to be
R $\sim$ 20\ 000 in the blue channel. 

Since \qt\ is intrinsically faint, we obtained 3 spectra of 30 minutes
each to minimize the unwanted effects of cosmic ray hits. After the
calibration was automatically produced \citep{Mittag2010}, we combined
the three spectra to increase the signal-to-noise ratio (S/N). We
estimated the final S/N of the combined spectra to be
\mbox{$\sim$40}. The date at which the HRT spectra was acquired is
indicated in Fig.~\ref{fig:follow_up} with a dashed vertical line.

\cite{herbst_1989} carried out a detail study on the equivalent width
(EW) of K- and M-type stars. These authors attempted to characterize
the activity level of the stars by producing a main sequence of EW as
a function of its color, representing the boundary between active and
low active stars relating two observables that are trivial to obtain,
i.e.,

\begin{equation}
  EW_{\mathrm{H}\alpha} = -1.49 + 1.95(R - I) - 0.77(R - I)^2\ ,
  \label{eq:EW}
\end{equation}

\begin{figure}[ht!]
  \centering
  \includegraphics[width=.5\textwidth]{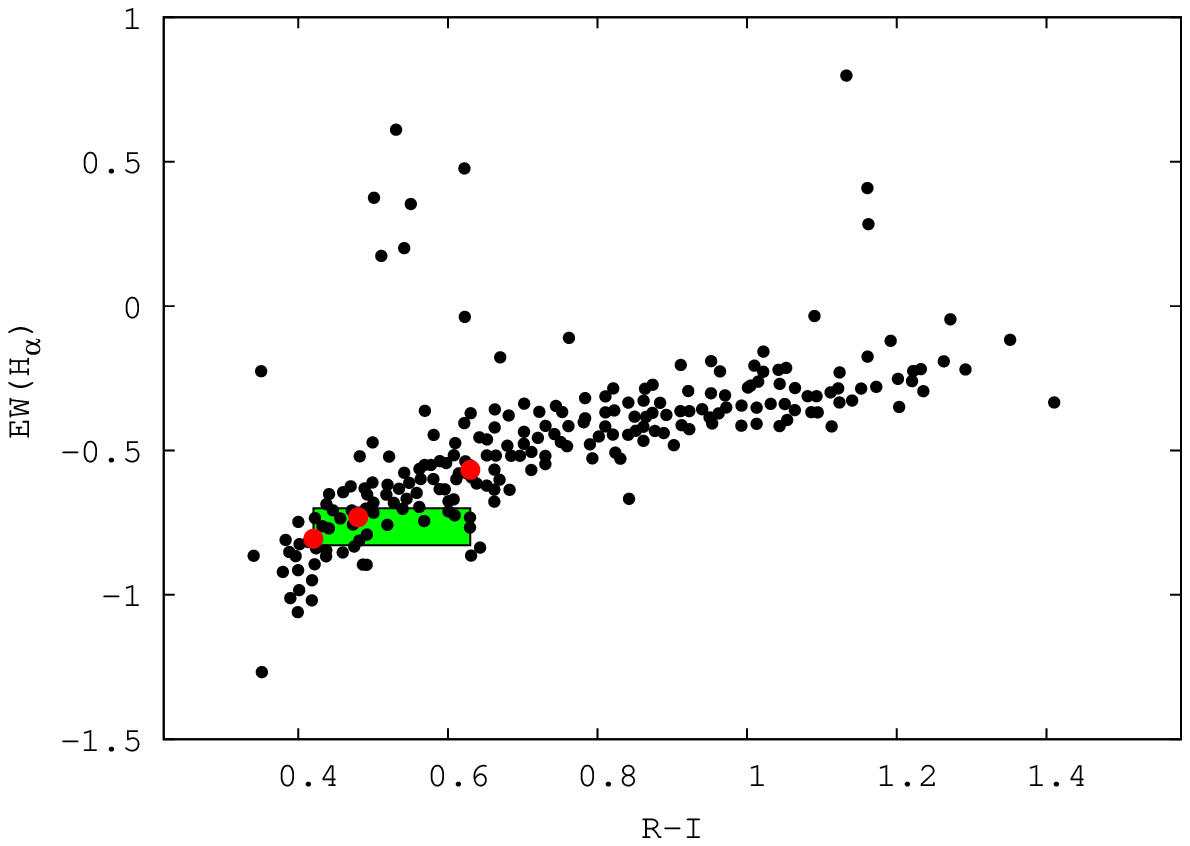}
  \caption{\label{fig:EW_RI}H$\alpha$ EW, as a function of R-I,
    obtained from \cite{herbst_1989} in black points, the values of
    $EW_{H\alpha}$ obtained using Eq.~\ref{eq:EW} from the same
    authors for three different color indexes in red points, and our
    estimated EW including the errors contained within the green
    rectangle.}
\end{figure}

\noindent The empirical relation is only valid for \mbox{R - I $>$
  0.4}. The spectral type of \mbox{Qatar-1} has been identified as K2V
\citep{Alsubai2011,Covino2013}. However, the reported color
information by \cite{droege_2006} (\mbox{V = 12.84 $\pm$ 0.14 mag},
\mbox{I = 11.71 $\pm$ 0.08}) is consistent with spectral types K0 to
K5. Following the Calibration of MK spectral types \citep{ALLEN}, K0
corresponds to \mbox{R - I = 0.42}, K2 to \mbox{R - I = 0.48}, and K5
to \mbox{R - I = 0.63}. Replacing these values into Eq.~\ref{eq:EW}
yields \mbox{$EW_{H\alpha} = -0.806$}, \mbox{$EW_{H\alpha} = -0.731$},
and \mbox{$EW_{H\alpha} = -0.567$}, respectively. Using the HRT
spectra we calculated the EW of the H$\alpha$ line in the usual way,
finding \mbox{$EW_{H\alpha,HRT} = -0.764 \pm
  0.064$}. Figure~\ref{fig:EW_RI} shows the EW--color index diagram of
the work of \cite{herbst_1989} (their Figure 4, black points), along
with the estimation of the EW of \mbox{Qatar-1} (red points) and the
calculated $EW_{H\alpha}$ values and their uncertainties (green
rectangle). The derived quantities place \mbox{Qatar-1} in the middle
of the distribution of the points of \cite{herbst_1989}. In other
words, \mbox{Qatar-1} falls over the boundary that divides active from
non-active stars. This is in agreement with the previously reported
moderate activity levels by \cite{Covino2013} and \cite{Mislis2015}.

\begin{acknowledgements}

C. von Essen acknowledges funding for the Stellar Astrophysics Centre,
provided by The Danish National Research Foundation (Grant DNRF106). A
special acknowledge to GEMINI Helpdesk, for replying so quickly and
diligently. S. Cellone acknowledges funding from UNLP through grant
11/G127, and thanks the Stellar Astrophysics Centre (Aarhus, Denmark)
for its hospitality. We acknowledge the referee for her/his positive
feedback.

\end{acknowledgements}

\bibliographystyle{aa}
\bibliography{vonEssenC}
\end{document}